\documentclass[12pt]{iopart}

\usepackage{iopams}
\usepackage{graphicx}
\usepackage{dcolumn}
\usepackage{bm}
\usepackage{float}
\newcommand{\bra}[1]{\langle#1|}

\newcommand{\ket}[1]{|#1\rangle}
\providecommand{\openone}{\leavevmode\hbox{\small1\kern-3.8pt\normalsize1}}
\begin{document}

\title{Entanglement dynamics in superconducting qubits affected by local bistable impurities}

\author{R. Lo Franco$^{1,3}$, A. D'Arrigo$^2$, G. Falci$^2$,  G. Compagno$^3$, and E. Paladino$^2$}
\address{$^1$Centro Siciliano di Fisica Nucleare e Struttura della Materia (CSFNSM) \& Dipartimento di Fisica e Astronomia, Universit\`a di Catania, Viale A. Doria 6, 95125 Catania, Italy\\
$^2$Dipartimento di Fisica e Astronomia, Universit\`a di Catania, viale A. Doria 6, 95125 Catania, Italy \& MATIS  CNR - INFM\\
$^3$CNISM \& Dipartimento di Fisica, Universit\`a di Palermo, via Archirafi 36, 90123 Palermo, Italy}
\ead{lofranco@fisica.unipa.it}

\begin{abstract}
We study the entanglement dynamics for two independent superconducting qubits each affected by a bistable impurity generating random telegraph noise (RTN) at pure dephasing. The relevant parameter is the ratio $g$ between qubit-RTN coupling strength and RTN switching rate, that captures the physics of the crossover between Markovian and non-Markovian features of the dynamics. For identical qubit-RTN subsystems, a threshold value $g_\mathrm{th}$ of the crossover parameter separates exponential decay and onset of revivals; different qualitative behaviors also show up by changing the initial conditions of the RTN. We moreover show that, for different qubit-RTN subsystems, when both qubits are very strongly coupled to the RTN an increase in entanglement revival amplitude may occur during the dynamics.
\end{abstract}



\maketitle

\section{\label{introduction}Introduction}
Entanglement dynamics has been intensively studied for bipartite quantum systems interacting with quantum environments (independent or common). It presents, depending on the Markovian or non-Markovian nature of the environments, phenomena like entanglement sudden death (ESD) \cite{yu2004PRL}, revivals \cite{bellomo2007PRL,mazzola2009PRA,Lopez2010PRA} or trapping \cite{bellomo2008trapping,bellomo2010PhyScr}. Entanglement dynamics has been also analyzed in some solid-state systems, like superconducting qubits or quantum dots, which are promising candidates for realizing quantum information processing \cite{imoto2002PRA,plenio2007NJP,palermocatania2010PRA,bellomo2011PhyScr}. A considerable progress has been made in the last decade towards the implementation of a controlled solid-state quantum computer. In particular, superconducting high-fidelity single qubit gates with coherence times of $\sim1 \mu$s are currently available \cite{vion,schreier}. High-fidelity Bell states of superconducting qubits have been also prepared \cite{steffen,schoelkopf2009Nature}. Superconducting nanodevices are usually affected by  broadband noise, with typical power spectra displaying a $1/f$ low-frequency behavior
followed by a white or ohmic flank \cite{ithier,nak-spectrum,paladino2002PRL,falci2005PRL}. Dynamics of entanglement in these devices has been recently analyzed distinguishing the effect of adiabatic and quantum noise \cite{palermocatania2010PRA}. The disentanglement process has been also studied by the quasi-Hamiltonian method when two independent superconducting qubits are locally affected by random telegraph noise (RTN) \cite{zhou2010QIP}.

In this paper we consider a model, relevant to solid-state system \cite{paladino2002PRL}, consisting of two noninteracting qubits each affected by a single bistable impurity inducing RTN at the pure dephasing working point. In addition, this model can be exactly solvable in analytic form and captures the physics of the crossover between a Markovian and non-Markovian environment under quite general conditions. The relevant parameter separating the two regimes is the ratio between qubit-RTN coupling constant and switching rate of the impurity. We analyze the entanglement dynamics both for identical and different conditions of the two local qubit-RTN. In Sec.~\ref{model} we introduce the relevant single qubit-RTN model, while in Sec.~\ref{dynent} we describe the entanglement dynamics under various noise initial conditions. In Sec.~\ref{conclusion}, we present our conclusive remarks.

\section{\label{model}Model: a qubit under pure dephasing RTN}
Our system consists of a couple of independent superconducting qubits, $A$ and $B$, each affected by a bistable impurity generating RTN at pure dephasing. The total Hamiltonian is $H_\mathrm{tot}=H_A+H_B$, where for each qubit the Hamiltonian at pure-dephasing is ($\hbar=1$) \cite{paladino2002PRL} $H=-(\Omega/2)\sigma_z-(v/2)\xi(t)\sigma_z$, where $\xi(t)$ is a stochastic process producing RTN switching at a rate $\gamma$ between $\pm1$ and $v$ is the qubit-RTN coupling constant. The power spectrum of the unperturbed equilibrium fluctuations of $\xi(t)$ is $s(\omega)=v^2\gamma/[2(\gamma^2+\omega^2)]$. The relevant parameter is the ratio $g=v/\gamma$ that permits to analyze the crossover between a Markovian environment for weakly coupled impurities ($g<1$) and a non-Markovian environment for strong coupled impurities ($g>1$) \cite{paladino2002PRL}. The exact evolution of single-qubit coherence $q(t)\equiv\rho_{01}(t)/\rho_{01}(0)$ has been found in Ref.~\cite{paladino2002PRL} and it is given by
\begin{equation}\label{singlequbitcoherence}
q(t)=\mathrm{e}^{-\mathrm{i}(\Omega+v/2)t}[A\mathrm{e}^{-\frac{\gamma(1-\alpha)t}{2}}
+(1-A)\mathrm{e}^{-\frac{\gamma(1+\alpha)t}{2}}],
\end{equation}
where $A=\frac{1}{2\alpha}(1+\alpha-\mathrm{i}g\delta p_0)$ and $\alpha=\sqrt{1-g^2}$. $\delta p_0$ is a degree of freedom depending on the initial conditions of the RTN and is not present in previous analysis of entanglement dynamics under RTN \cite{zhou2010QIP}. In particular, $\delta p_0$ can take the value $\delta p_0=0$ (corresponding to the value of  thermodynamical equilibrium) or the values $\delta p_0=\pm1$. The form of Eq.~(\ref{singlequbitcoherence}) clearly shows the different roles of weakly and strongly coupled impurities in the decoherence process. Dephasing comes from the
sum of two exponential terms. If $g\ll1$ only the first of these terms is important and the corresponding rate is $\approx v^2/(4\gamma)$, coinciding with the golden rule; if $g\gg1$ the two terms are of the same order and the decay rate is $\sim\gamma$ \cite{paladino2002PRL}.

\section{Dynamics of entanglement\label{dynent}}
The two-qubit density matrix elements are evaluated in the computational basis $\mathcal{B}=\{\ket{0}\equiv\ket{00},\ket{1}\equiv\ket{01},\ket{2}\equiv\ket{10},\ket{3}\equiv\ket{11}\}$, where $H_{i}\ket{0_i}=-\frac{\Omega_i}{2}\ket{0_i}$, $H_{i}\ket{1_i}=\frac{\Omega_i}{2}\ket{1_i}$ ($i=A,B$). We consider as initial states the extended Werner-like (EWL) states expressed by the density matrices \cite{bellomo2008PRA}
\begin{equation}\label{EWLstates}
    \hat{\rho}_1=r \ket{1_{a}}\bra{1_{a}}+\frac{1-r}{4}\openone_4,\quad
    \hat{\rho}_2=r \ket{2_{a}}\bra{2_{a}}+\frac{1-r}{4}\openone_4,
\end{equation}
whose pure parts are the one-excitation and two-excitation Bell-like states $\ket{1_{a}}=a\ket{01}+b\ket{10}$, $\ket{2_{a}}=a\ket{00}+b\ket{11}$, where the subscript $a$ identifies the initial degree of entanglement of the pure part and $|a|^2+|b|^2=1$. The density matrix of EWL states is non-vanishing only along the diagonal and anti-diagonal (X form) \cite{bellomo2008PRA} and this structure is maintained at $t>0$ in the system we are considering (pure dephasing). Using the concurrence \cite{wootters1998PRL} $C$ to quantify entanglement, the initial entanglement is equal for both the EWL states of Eq.~(\ref{EWLstates}) and reads $C_{\rho_1}(0)=C_{\rho_2}(0)=2\mathrm{max}\{0,(|ab|+1/4)r-1/4\}$. Initial states are thus entangled for $r>r^\ast=(1+4|ab|)^{-1}$. Moreover, the purity $P=\mathrm{Tr}(\rho^2)$ of EWL states is $P=(1+3r^2)/4$. Entangled states with purity $\approx 0.87$ and fidelity to ideal Bell states $\approx 0.90$ have been experimentally generated \cite{schoelkopf2009Nature}: these states may be approximately described as EWL states with $r_\mathrm{exp}\approx0.91$.

In order to obtain the concurrence at time $t$, we need the evolved two-qubit density matrix which can be evaluated by the knowledge of the single-qubit density matrix evolution, according to a standard procedure \cite{bellomo2007PRL}. The initial EWL states, during the pure-dephasing evolution, maintain the diagonal elements unchanged and the anti-diagonal elements depend on product of single-qubit coherences. The concurrences at time $t$ for the two initial states of Eq.~(\ref{EWLstates}) are given by, respectively, $C_{\rho_1}(t)=2\mathrm{max}\{0,K_1(t)\}$ and $C_{\rho_2}(t)=2\mathrm{max}\{0,K_2(t)\}$ where $K_1(t)=|\rho_{12}(t)|-\sqrt{\rho_{00}(t)\rho_{33}(t)}$,
$K_2(t)=|\rho_{03}(t)|-\sqrt{\rho_{11}(t)\rho_{22}(t)}$. In our case we have $K_1(t)=K_2(t)=K(t)$, so that the two concurrences are equal for both initial states $C_{\rho_1}(t)=C_{\rho_2}(t)=C(t)$. In the following, we consider identical qubits ($\Omega_A=\Omega_B=\Omega$) and we distinguish the cases of equal RTNs $g_A=g_B=g$ and of different RTNs $g_A\neq g_B$ for the two subsystems.

\subsection{Identical RTNs}
\begin{figure}
\begin{center}
{\includegraphics[width=0.45\textwidth]{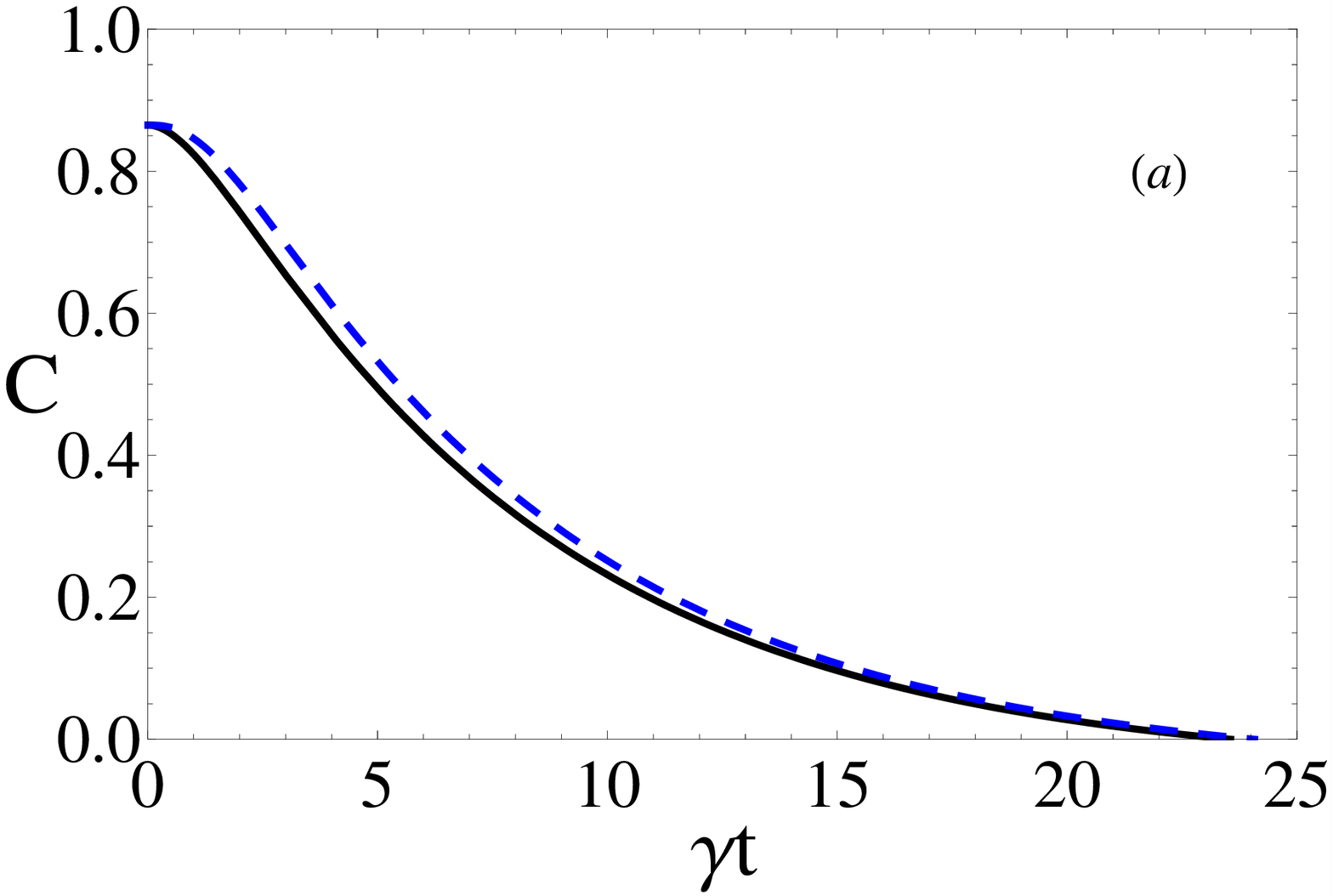}
\hspace{0.5 cm}
\includegraphics[width=0.45\textwidth]{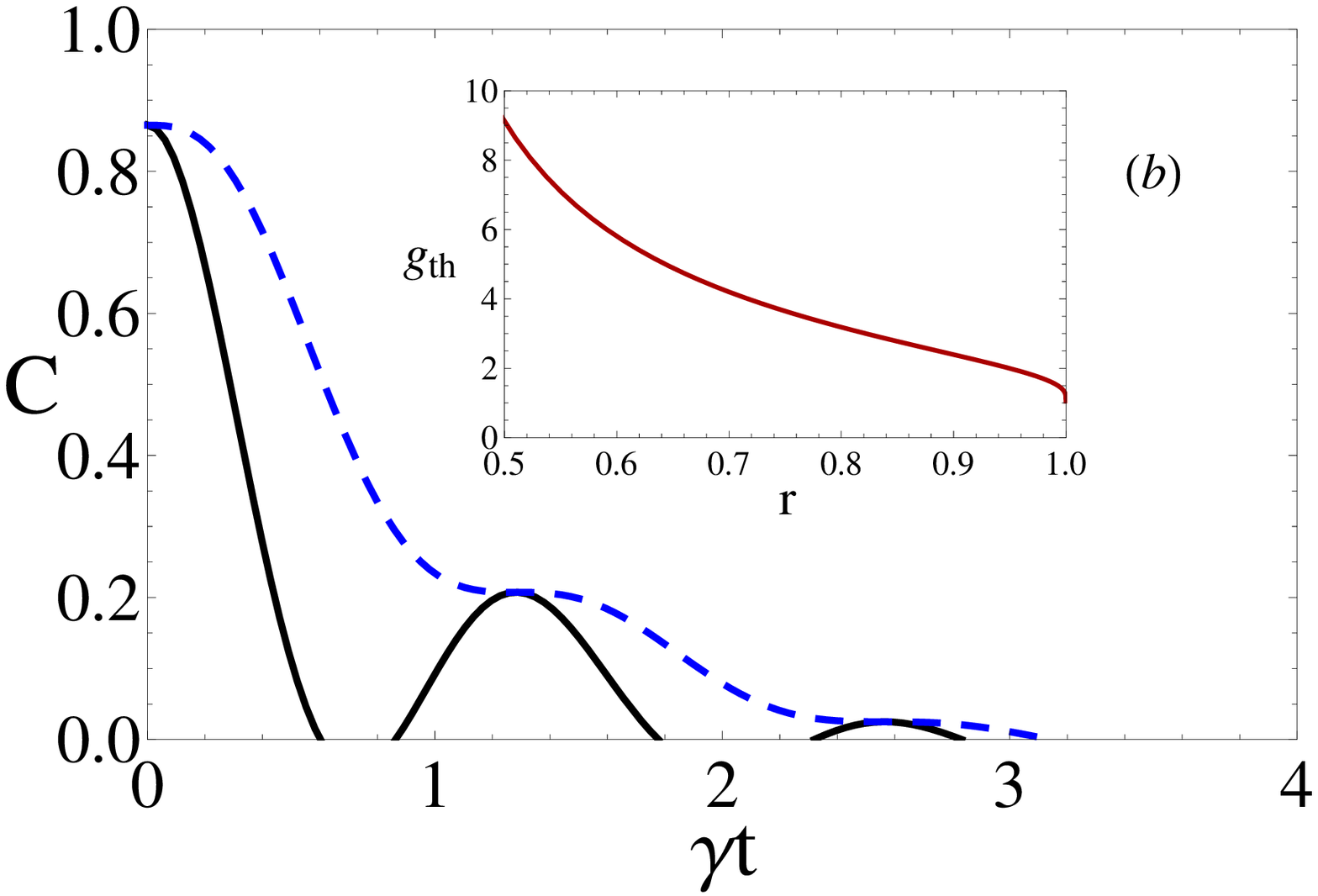}}
\caption{\label{fig:concurrence1}\footnotesize Concurrences versus the dimensionless time $\gamma t$ with $r=r_\mathrm{exp}=0.91$ and $a=b=1/\sqrt{2}$ for $g=0.5$ (panel (a)) and $g=5$ (panel (b)). Solid black lines are for $\delta p_0=0$ while dashed blue lines for $\delta p_0=\pm1$. The inset of panel (b) displays $g_\mathrm{th}$ versus $r$ for $a=b=1/\sqrt{2}$.}
\end{center}
\end{figure}

Consider the case $g_A=g_B=g$ and equal initial conditions. We obtain $C(t)=\mathrm{max}\{0,2K(t)\}$ with
\begin{equation}\label{Kt}
K(t)=r|a|\sqrt{1-|a|^2}|q(t)|^2-(1-r)/4,
\end{equation}
where $q(t)$ is the single-qubit coherence of Eq.~(\ref{singlequbitcoherence}). Now we investigate how the concurrence depends on the different initial conditions of the system and state parameters. Equation (\ref{Kt}) clearly shows the role of the purity parameter $r$ on concurrence evolution: for pure states, $r=1$, there cannot be an entanglement sudden death (ESD) and the entanglement goes asymptotically to zero similarly to single-qubit coherence; on the other hand, any value of $r$ inside the interval $r^\ast<r<1$ determines a threshold value for the first term of Eq.~(\ref{Kt}) to be overcome in order to have nonzero entanglement (of course, $0<|a|<1$). The dynamics of entanglement is shown in Fig.~\ref{fig:concurrence1}, when $g=0.5,5$ and $\delta p_0=0,\pm1$, for initial states with $r=r_\mathrm{exp}=0.91$ and $a=b=1/\sqrt{2}$ whose initial concurrence is $C(0)=0.865$.
Similarly to other works \cite{bellomo2007PRL,zhou2010QIP}, for weak coupling ($g<1$, Markovian environment) there is a simple (exponential) decay while for strong coupling ($g>1$, non-Markovian environment) there are damped revivals after dark periods of entanglement (see Fig.~\ref{fig:concurrence1}). It is however worth to note that, while for a single qubit the crossover between Markovian (exponential) and non-Markovian (oscillating) decay is identified by $g=1$, for the two-qubit system, depending on initial state parameters, a threshold value $g_\mathrm{th}>1$ exists separating exponential decay and the onset of revivals. This $g_\mathrm{th}$ can be found analytically \cite{lofrancoinprogress} and its dependence on $r$ (for $a=b=1/\sqrt{2}$) is displayed in Fig.~\ref{fig:concurrence1}(b) (for $r=1$ it is $g_\mathrm{th}=1$). When $r=1$ the entanglement goes to zero asymptotically (oscillating and vanishing at given times for $g>g_\mathrm{th}$), analogously with the results already found for the case of adiabatic (low-frequency) noise due to collective impurities generating $1/f$-noise \cite{palermocatania2010PRA}. Differently, when $r<1$ there is always ESD for $g\leq g_\mathrm{th}$ and a ``final death'' (after revivals) for $g>g_\mathrm{th}$, where for final death we mean the definitive disappearance of entanglement. Different RTN initial conditions (values of $\delta p_0$) qualitatively affects the entanglement dynamics for $g>g_\mathrm{th}$ (strong memory effects), while leave it practically unchanged for $g\leq g_\mathrm{th}$ (weak memory effects), as shown in Fig.~\ref{fig:concurrence1}. In particular, when $g>g_\mathrm{th}$, the concurrence for $\delta p_0=\pm1$ is always larger than that for $\delta p_0=0$ with small beats touching the peaks of the revivals appearing in the corresponding curve for $\delta p_0=0$; the final death time is also longer than the previous one (see Fig.~\ref{fig:concurrence1}).

\begin{figure}
\begin{center}
\includegraphics[width=0.45\textwidth]{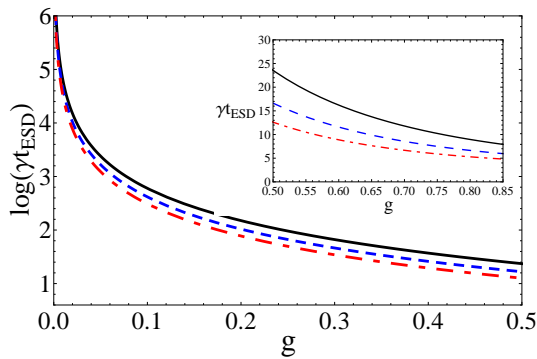}
\caption{\label{fig:tESD}\footnotesize ESD times $t_\mathrm{ESD}$ (scaled with $\gamma$) versus $g$, with $a=b=1/\sqrt{2}$ and $\delta p_0=0$, for $r_\mathrm{exp}=0.91$ (black solid line), $r=0.8$ (blue dashed line) and $r=0.7$ (red dot-dashed line). The inset shows the range $g>0.5$.}
\end{center}
\end{figure}
In the limit of small $g$ (Markovian noise) and for $\delta p_0=0$, only the first term is important in Eq.~(\ref{singlequbitcoherence}) and, from Eq.~(\ref{Kt}), we may estimate the ESD times (giving however good match with plots up to $g=0.9$)
\begin{equation}\label{ESDtimes}
\gamma t_\mathrm{ESD}=-\frac{2}{1-\sqrt{1-g^2}}\ln\left(\frac{\sqrt{\frac{(1-g^2)(1-r)}{r|a| \sqrt{1-|a|^2}}}}{1+\sqrt{1-g^2}}\right).
\end{equation}
These ESD times are plotted in Fig.~\ref{fig:tESD} as a function of $g$ for three different values of $r$. They decrease as $g$ increases, with a reduction of about an order of magnitude going from $g=0.1$ to $g=0.4$.

\subsection{Different RTNs}
\begin{figure}[h]
\begin{center}
{\includegraphics[width=0.45\textwidth]{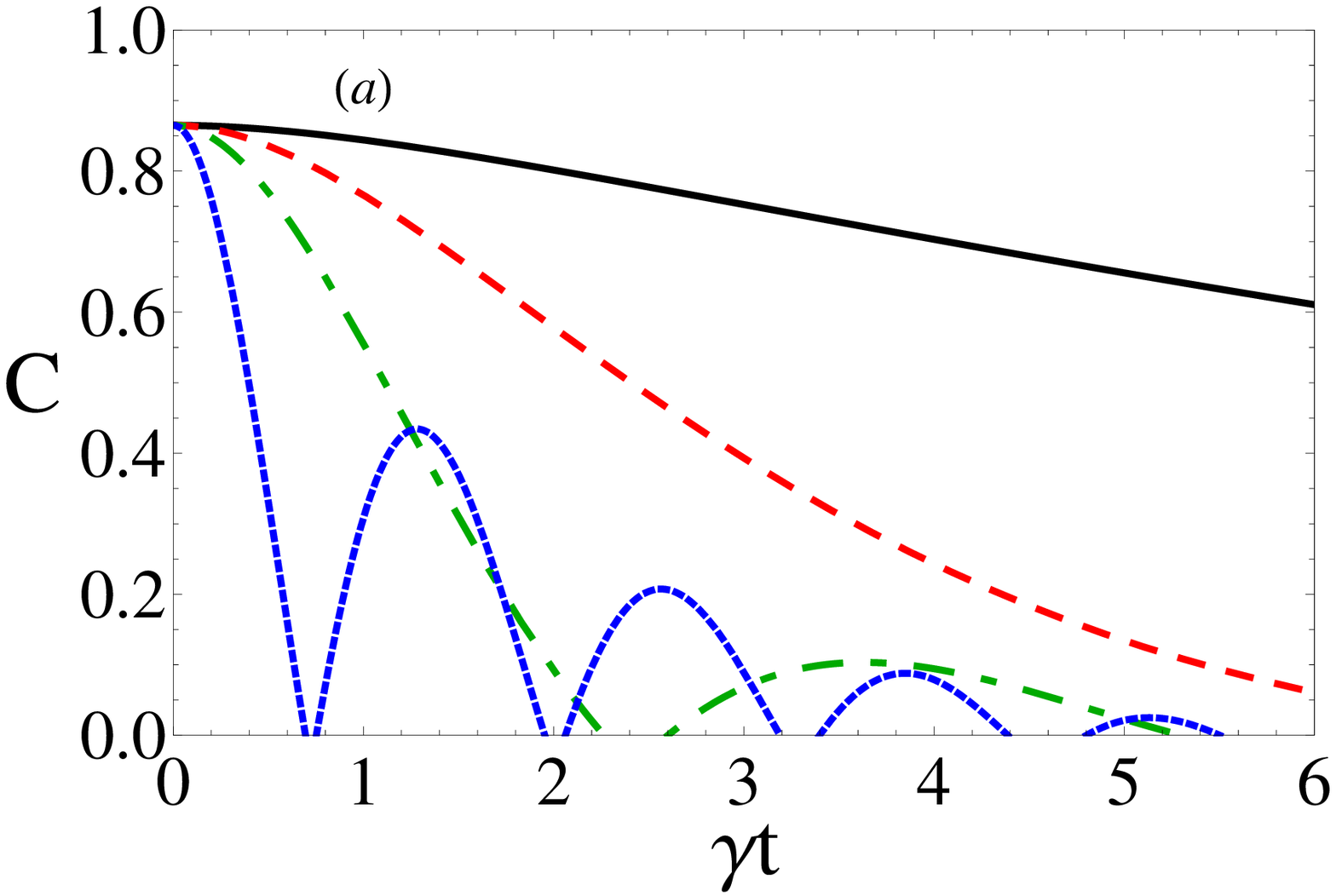}
\hspace{0.5 cm}
\includegraphics[width=0.45\textwidth]{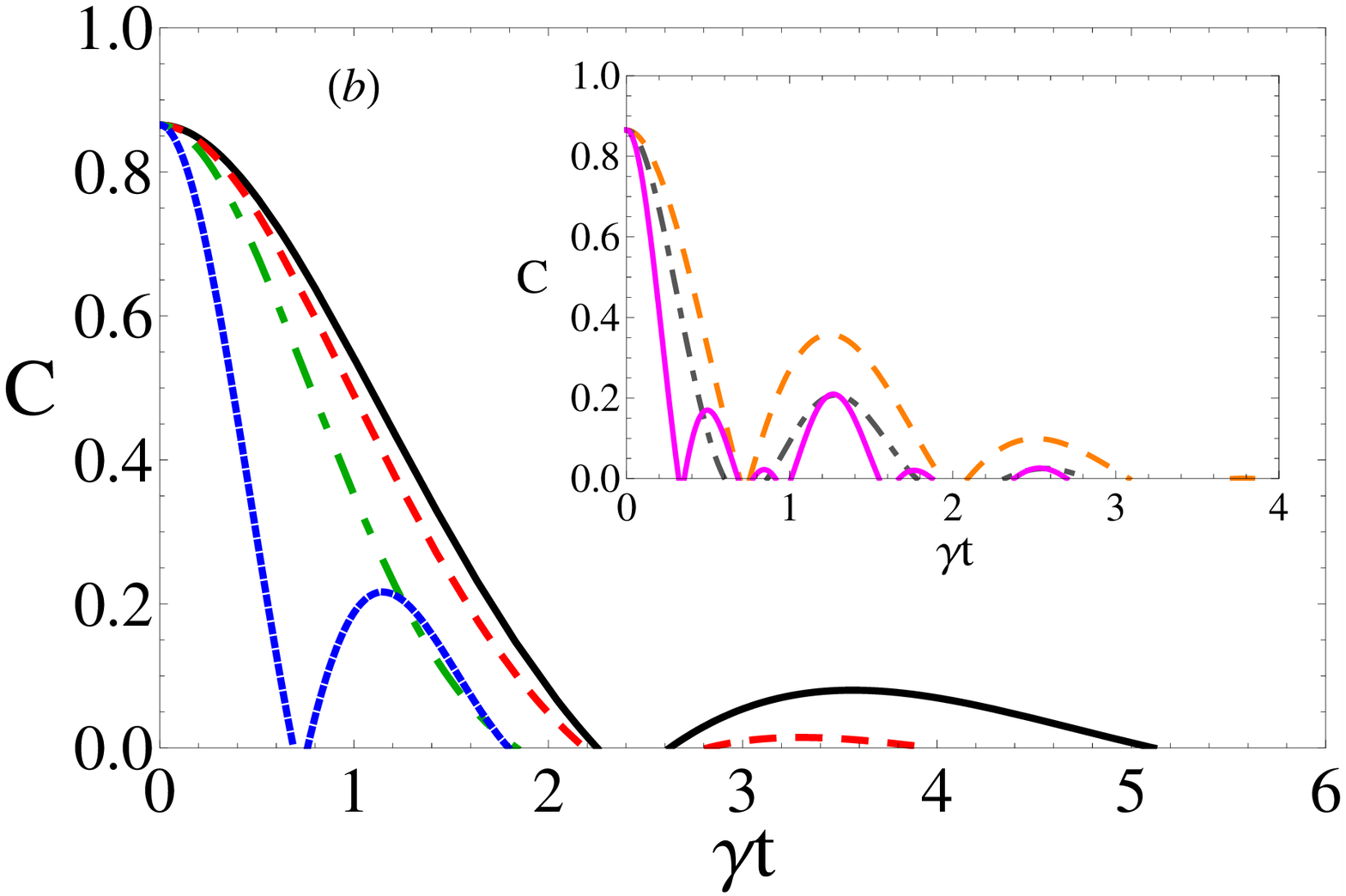}}
\caption{\label{fig:concurrencedifferentenv}\footnotesize Concurrences versus $\gamma t$ for $g_B=0$ (panel (a)) and $g_B=2$ (panel (b)) and values of $g_A$ equal to: $0.5$ (solid black line), $1.1$ (red dashed line), $2$ (green dot-dashed line), $5$ (blue dotted line). In the inset of panel (b) we set $g_B=5$ with values of $g_A$ equal to: $1.1$ (orange dashed line); $5$ (gray dot-dashed line); $10$ (magenta solid line). Initial conditions: $r=r_\mathrm{exp}=0.91$, $a=b=1/\sqrt{2}$, $\delta p_0=0$.}
\end{center}
\end{figure}

We now consider the more realistic case of qubits affected by different RTNs $g_A\neq g_B$, with $\delta p_0=0$ for both. The function $K(t)$ is now given by Eq.~(\ref{Kt}) replacing $|q(t)|^2$ with $|q_A(t)q_B(t)|$, where $q_i(t)$ ($i=A,B$) is the coherence of qubit $i$ given in Eq.~(\ref{singlequbitcoherence}). The entanglement dynamics for different values of $g_A,g_B$ is displayed in Fig.~\ref{fig:concurrencedifferentenv}. We assume the rate $\gamma$ fixed for both impurities and different values of $v_A,v_B$.
When a qubit, for instance qubit $B$, is not affected by RTN ($g_B=0$) we find a qualitative behavior analogous to that observed for identical RTNs, with a threshold value $g_{A\mathrm{th}}>1$ after which revivals occur. On the other hand, when one of the two qubits is affected by a strong non-Markovian RTN (for instance, $g_B=2$) when $g_A$ increases the revivals tend to disappear after a certain $g_A>1$ (indeed, when $g_A=g_B=2$ the value of $g_\mathrm{th}$ is larger than 2, as seen in previous section) and then reappear for larger values of $g_A$ with final death times shorter and shorter. By further increasing the fixed value of $g_B$, other new qualitative behaviors appear when $g_A$ increases in the strong coupling domain. Inset of Fig.~\ref{fig:concurrencedifferentenv}(b) shows the presence of entanglement revivals whose amplitude increases with respect to previous one for the values $g_A=10,g_B=5$. Similar behaviors occur for initial pure states ($r=1$) but without dark periods. These behaviors are due to the different contributions of the two single-qubit dynamics.

\section{Conclusions \label{conclusion}}
In this work we have analyzed the entanglement dynamics in a system of two initially entangled independent superconducting qubits ($A$ and $B$) each affected by impurity-induced random telegraph noise at pure dephasing. A crucial role in determining the behavior of entanglement dynamics is played by the ratio between qubit-RTN coupling $v$ and switching rate $\gamma$, $g=v/\gamma$. For identical RTNs we have found that, in spite of the fact that for the single-qubit dynamics the crossover between Markovian and non-Markovian behavior is identified by $g=1$, for an initially entangled two-qubit system this threshold is in general shifted to a value $g_\mathrm{th}>1$ depending on the initial state (for initial pure states $g_\mathrm{th}=1$). In the case when the RTN has the initial value $\delta p_0=0$ (thermodynamical equilibrium), we have retrieved the known behaviors of exponential (Markovian) decay for $g\leq g_\mathrm{th}$ and of oscillating (non-Markovian) decay with revivals for $g>g_\mathrm{th}$. We have then found new qualitative behaviors, in the non-Markovian regime ($g>g_\mathrm{th}$, relevant memory effects), by changing the initial value of RTN, $\delta p_0$. For example, if $\delta p_0=\pm1$, the entanglement never vanishes at intermediate times and no revivals occur, with a final death time longer than that for $\delta p_0=0$. We have finally considered the case when the two local qubit-RTN conditions differ, in particular $g_A\neq g_B$. Entanglement dynamics may behave quite differently than that for identical subsystems. In particular, when both $g_A,g_B$ are in the strong non-Markovian domain a new feature in the entanglement dynamics shows up, that is there can be entanglement revivals whose amplitude increases with respect to previous one.

The simple model, relevant to solid-state systems, analytically studied here displays the richness of behaviors of  entanglement dynamics and its crucial dependence both on system-environment parameters and on initial conditions.

\section*{Acknowledgments}
Partially supported by EU through Grant No. PITN-GA-2009-234970.

\section*{References}

\bibliography{bibRTN}

\end{document}